\documentclass{article}

\title{Relativistic Ladder Operators for the Three-Dimensional Harmonic Oscillator}
\author{Robert Ducharme}

\begin{document}
\maketitle

\centerline{151 Fairhills Dr., Ypsilanti, MI 48197}
\centerline{E-mail: robert@faac.com}

\begin{abstract}
The quantum constraint equations for a relativistic three-dimensional harmonic oscillator are shown to find concise expression in terms of Lorentz covariant ladder operators. These ladder operators consist of two conjugate 4-vectors that are each constrained to generate three linearly independent combinations of ladder operator components for raising and lowering the eigenstates of the oscillator. Correspondence to the Schr\"{o}dinger equation for the harmonic oscillator in the non-relativistic limit is demonstrated.
\end{abstract}

\section{Introduction}
Quantum Constraint Mechanics (QCM) is a generalized form of Hamiltonian quantum mechanics that enables bound systems of interacting particles to be described in a manifestly covariant manner \cite{AK, RST, CA}. QCM has been applied to problems in nuclear \cite{LC,WC} and high-energy physics \cite{CWA}. The purpose of this paper to show the quantum constraint equations for a relativistic two-particle harmonic oscillator having a total 4-momentum $P_{\mu}$ $(\mu=1,2,3,4)$ and total mass $M_0$ have a concise expression in terms of Lorentz covariant raising $\hat{a}_\mu^{+}$ and lowering $\hat{a}_\mu^{-}$ operators. The forms of these relativistic ladder operators are determined directly from the assumption that they reduce to the form $\hat{a}_\mu^{\pm} = (\hat{a}_i^{\pm}, 0)$ in the rest frame of the oscillator where $\hat{a}_i^{\pm}$ are the non-relativistic ladder operators.

The quantum constraint equations for the relativistic 3-dimensional harmonic oscillator are presented in section 2 of this paper alongside the solution $\Psi$ . Explicit expressions for the Lorentz covariant ladder operators are derived in section 3. It is shown, in particular, that the set of quantum constraints on $\Psi$ can be written in the compact form $\hat{P}_\mu \hat{P}^\mu \Psi=  -M_0^2 \Psi$, $\hat{a}_\mu^{+} \hat{a}^{\mu-} \Psi = n \Psi$ and $\hat{P}^\mu \hat{a}_\mu^{\pm} \Psi = 0$ where $(n+\frac{3}{2})\Omega$ is the total (energy)$^2$ of the oscillator and $\Omega$ is the spring constant.

In non-relativistic quantum mechanics there is one component of each of the $\hat{a}_i^{\pm}$ ladder operators for each orthogonal mode of oscillation. By contrast, in the relativistic approach developed here, the $\hat{P}^\mu \hat{a}_\mu^{\pm} \Psi = 0$ constraint condition reduces the number of independent components of $\hat{a}_\mu^{\pm}$ from four to three. The three operators required for raising and lowering the eigenstate of the oscillator are therefore linear combinations of $\hat{a}_\mu^{\pm}$ components. Explicit forms for these linear combination operators are derived in section 4. 

The 4-vector convention in this paper will be $x_\mu=(x_1, x_2, x_3, x_4)$ and $x^\mu=(x_1, x_2, x_3, -x_4)$. Natural units, $c=\hbar=1$ will be used throughout. 

\section{The Constraint Equations}
Consider a 3-dimensional oscillator consisting of two interacting particles each to be represented using an index $k(=1,2)$.  Let $m_{k}$, $x_{k\mu}$ and $p_{k\mu}$ denote the mass, 4-position and 4-momentum of each of the particles respectively. Following Crater and Van Alstine \cite{CA} the center-of-mass and relative coordinates for the system can then be expressed as
\begin{equation} \label{eq: Xdef}
X_{\mu} = M_0^{-1}(\epsilon_1 x_{1\mu} + \epsilon_2 x_{2\mu}), \quad P_{\mu} = p_{1\mu} + p_{2\mu}
\end{equation}
\begin{equation} \label{eq: xdef}
x_{\mu} = x_{1\mu} - x_{2\mu}, \quad p_{\mu} = M_0^{-1}(\epsilon_2 p_{1\mu} - \epsilon_1 p_{2\mu})
\end{equation}
where the $\epsilon_1$ and $\epsilon_2$ parameters take the form
\begin{equation} \label{eq: epsilon1}
\epsilon_1 = (2M_0)^{-1}(M_0^2+m_1^2-m_2^2)
\end{equation}
\begin{equation} \label{eq: epsilon2}
\epsilon_2 = (2M_0)^{-1}(M_0^2+m_2^2-m_1^2)
\end{equation}
The operators for 4-momentum in these two coordinate systems are
\begin{equation} \label{eq: qmreps}
\hat{p}_{k\mu}=\frac{1}{i}\frac{\partial}{\partial x_k^{\mu}}, \quad
\hat{P}_{\mu}=\frac{1}{i}\frac{\partial}{\partial X^{\mu}}, \quad
\hat{p}_{\mu}=\frac{1}{i}\frac{\partial}{\partial x^{\mu}}
\end{equation}
It will be further assumed that the oscillator is in a free state such that the total momentum $P_{\mu}$ satisfies the relationship
\begin{equation} \label{eq: totMom}
P^\mu P_\mu = -M_0^2
\end{equation}
where $M_0$ is the total mass of the oscillator. 

It is usual in QCM, to further define the internal constraint space coordinates in both 4-position space
\begin{equation} \label{eq: xperp}
x_{\perp}^\mu = x^\mu + M_0^{-2}P^\mu (P_\nu x^\nu)
\end{equation}
and 4-momentum space
\begin{equation} \label{eq: pperp}
p_{\perp}^\mu = p^\mu + M_0^{-2} P^\mu (P_\nu p^\nu)
\end{equation}
Pre-multiplying these results by $P_\mu$ gives
\begin{equation} \label{eq: xcon}
P_\mu x_{\perp}^\mu = 0
\end{equation}
\begin{equation} \label{eq: pcon}
P_\mu p_{\perp}^\mu = 0
\end{equation}
showing $x_{\perp}^\mu = (x^1,x^2,x^3,0)$ and $p_{\perp}^\mu = (p^1,p^2,p^3,0)$ in the rest frame of the oscillator. 

The scalar potential (energy)$^2$ term for two interacting particles in a harmonic oscillator can be written $\Omega^2 x_\perp^2$ where $\Omega$ is the spring constant. The relativistic wave equation constraining the wavefunction $\Psi(x_\mu, X_\mu)$ for two particles interacting through this oscillator potential can be expressed as
\begin{equation}
\hat{K}_T \Psi = (\hat{p}_1^2+\hat{p}_2^2+m_1^2+m_2^2+4\Omega^2 x_\perp^2)\Psi = 0
\end{equation}
This clearly reduces to the sum of two free-particle Klein-Gordon equations in the case $\Omega = 0$. Use of eqs. (\ref{eq: Xdef}) and (\ref{eq: xdef}) enables the $\hat{K}_T$ operator to be transformed into center-of-mass and relative coordinates giving
\begin{equation} \label{eq: TotHamil}
\hat{K}_T \Psi = [\hat{P}^2+m_c^2+4(\hat{p}^2 + \Omega^2 x_\perp^2)]\Psi = 0 
\end{equation}
where $m_c=m_1+m_2$ is the combined mass of the two particles. A quantum system of two particles implies the existence of two first-order constraints. For the second condition, Crater and Van Alstine \cite{CA} find
\begin{equation} \label{eq: SubCondition}
\hat{K}_S\Psi = P_{\mu} \hat{p}^{\mu} \Psi = 0
\end{equation}
Here, the operators for $\hat{K}_T$ and $\hat{K}_S$ commute meaning that energy is conserved as the system evolves in a single time parameter. 

Eqs. (\ref{eq: TotHamil}) and (\ref{eq: SubCondition}) have a separable solution of the form $\Phi(x_\mu) \exp(iP_\mu X^\mu)$. Use of eq. (\ref{eq: totMom}) to simplify the $K_T \Psi = 0$ constraint condition gives the internal harmonic oscillator equation
\begin{equation} \label{eq: HOE}
(\hat{p}^2+\Omega^2  x_\perp^2) \Psi = 2\sigma \Psi
\end{equation}
where the rest mass of the oscillator can be determined from the condition
\begin{equation} \label{eq: totMass}
M_0^2 = m_1^2+m_2^2+4\sigma + \sqrt{[(m_1^2+m_2^2+4\sigma)^2 - (m_1^2-m_2^2)^2]}
\end{equation}
It can be seen that eq. (\ref{eq: totMass}) gives $M_0=m_c$ for the free-particle case $(\sigma=0)$ and $M_0^2=m_c^2+8\sigma$ for $m_1=m_2$.  The non-relativistic limit of this expression is discussed at the end of this section.

Eq. (\ref{eq: HOE}) is readily solved to give the internal oscillator function
\begin{eqnarray} \label{eq: psi} 
\Phi(x_\mu) = \frac{1}{\sqrt{2^n}} \left( \frac{\Omega}{\pi} \right)^{3/4} \prod_{i=1}^{i=3} \left[\frac{1}{\sqrt{l_i!}} H_{l_i}(\sqrt{\Omega} x_i^{\prime}) \exp \left(-\frac{\Omega x_{i}^{\prime 2}}{2} \right)\right]
\end{eqnarray}
In this, $H_{l_i}$ are Hermite polynomials and $l_i=0,1,2...$. Inserting eq. (\ref{eq: psi}) into (\ref{eq: HOE}) gives
\begin{equation} \label{eq: sigma}
\sigma = \Omega\left(\frac{3}{2} + n\right)
\end{equation}
where $n=l_1 + l_2 + l_3$. 

Eq. (\ref{eq: psi}) preserves its form under the general Lorentz transformation
\begin{equation}
x_i^{\prime} = x_i + \gamma v_i \left(\frac{\gamma v_j x_j }{1+\gamma} - t \right)
\end{equation}
\begin{equation}
t^{\prime} = \gamma(t-v_j x_j)
\end{equation}
(see ref. \cite{CA}) having set $x_\mu=(x_i,t)$. Here, $v_i$ is the velocity of the oscillator and $\gamma = (1-v^2)^{-1/2}$.

The oscillator function (\ref{eq: psi}) has been normalized using the condition
\begin{equation}
\int \Psi^\dag \Psi \delta(M_0^{-1}P_\mu x^\mu)d^4x = 1
\end{equation}
where $\delta(M_0^{-1}P_\mu x^\mu)$ is the Dirac delta function restricting the oscillator to a 3-dimensional constraint space. Here, the choice of constraint space is purely for convenience and simply represents one possible choice of consistent convention of how observers in different inertial reference frames make measurements.  

In the non-relativistic limit $(v\ll 1, \sigma \ll m_c^2)$ , the wavefunction $\Psi$ loses its dependence on the relative time t such that eq. (\ref{eq: SubCondition}) reduces to the form $\partial \Psi / \partial t \simeq 0$. Also, the relativistic rest mass $M_0$ given in eq. (\ref{eq: totMass}) can be approximated to give $M_0  \simeq m_c + E_K$ where $E_K = \sigma / m_r$ is the internal kinetic energy of the oscillator and $m_r = m_1m_2 / m_c$ is the reduced mass of the particles. Inserting these results into eq. (\ref{eq: HOE}) leads to the Schr\"{o}dinger equation for the 2-particle harmonic oscillator:
\begin{equation} \label{eq: shrod1}
\frac{1}{2 m_r} \frac{\partial^2 \Psi}{\partial x_i \partial x_i} + \frac{1}{2} m_r \omega^2 x_ix_i\Psi = E_K\Psi
\end{equation}
where $\omega=\Omega/ m_r$. 

\section{Ladder Operators}
As is well known \cite{DFL}, the Schr\"{o}dinger equation (\ref{eq: shrod1}) can be simplified in terms of the non-relativistic raising and lowering operators: 
\begin{equation} \label{eq: ladder_nr}
\hat{a}_i^{\pm}(\hat{x}_i,\hat{p}_i) = \frac{1}{\sqrt{2 \Omega}} \left( \mp i \hat{p}_i + \Omega \hat{x}_i \right)
\end{equation}
to give 
\begin{equation} \label{eq: shrod2}
\left(\hat{a}_i^+ \hat{a}_i^-+\frac{3}{2} \right)\omega \Psi = E_K \Psi
\end{equation}
Eq. (\ref{eq: shrod2}) can also be written in the following equivalent form
\begin{equation} \label{eq: shrod3}
\left(\hat{a}_i^+ \hat{a}_i^-+\frac{3}{2} \right)\Omega \Psi = \sigma \Psi
\end{equation}
or even more concisely as
\begin{equation} \label{eq: shrod4}
\hat{a}_i^+ \hat{a}_i^- \Psi = n \Psi
\end{equation}
Eq. (\ref{eq: xperp}) suggests a Lorentz covariant generalization of eq. (\ref{eq: ladder_nr}) to be
\begin{equation} \label{eq: ladder_lc}
\hat{a}^{\mu \pm}(\hat{x}_{\perp}^\mu,\hat{p}^\mu) = \frac{1}{\sqrt{2 \Omega}} \left( \mp i \hat{p}^\mu + \Omega \hat{x}_{\perp}^{\mu} \right)
\end{equation}
In comparing, eqs. (\ref{eq: ladder_nr}) and (\ref{eq: ladder_lc}) it can be seen $\hat{x}_{\perp}^\mu$ replaces $\hat{x}^\mu$ but that $\hat{p}_{\perp}^\mu$ does not need to replace  $\hat{p}^\mu$ since eqs. (\ref{eq: pperp}) and (\ref{eq: SubCondition}) give $\hat{p}_{\perp}^\mu=\hat{p}^\mu$. It follows therefore that eq. (\ref{eq: ladder_lc}) reduces exactly to eq. (\ref{eq: ladder_nr}) in the rest frame of the oscillator.

Evaluating the product $\hat{a}_\mu^{+}\hat{a}^{\mu-}$ using eq. (\ref{eq: ladder_lc}) gives
\begin{equation}\label{eq: aprop1}
\hat{a}_\mu^{+}\hat{a}^{\mu-}  = (2 \Omega)^{-1}(\hat{p}^2+\Omega^2 x_\perp^2 - 3 \Omega)
\end{equation}
This expression can be rewritten as
\begin{equation}\label{eq: aprop2}
\left(\hat{a}_{\mu}^+ \hat{a}^{\mu-}+\frac{3}{2}\right)\Omega \Psi = \sigma \Psi
\end{equation}
having used eqs. (\ref{eq: HOE}) and (\ref{eq: sigma}). Here, it can be seen eqs. (\ref{eq: shrod3}) and (\ref{eq: aprop2}) are identical in form except in passing from the non-relativistic to the relativistic forms that $\mu$ replaces i as the summation index.

Hence, using eqs. (\ref{eq: xcon}), (\ref{eq: ladder_lc}) and (\ref{eq: aprop2}) to simplify the constraint eqs. (\ref{eq: TotHamil}) and (\ref{eq: SubCondition}) leads to
\begin{equation} \label{eq: KG}
\hat{P}_\mu \hat{P}^\mu \Psi=  -M_0^2 \Psi
\end{equation}
\begin{equation} \label{eq: lo1}
\hat{a}_\mu^{+} \hat{a}^{\mu-} \Psi = n \Psi
\end{equation}
\begin{equation} \label{eq: lo2}
\hat{P}^\mu \hat{a}_\mu^{\pm} \Psi = 0
\end{equation}
Eqs. (\ref{eq: KG}) through (\ref{eq: lo2}) constitute a complete set of quantum constraint equations for the relativistic 3-dimensional harmonic oscillator expressed in terms of the Lorentz covariant ladder operators (\ref{eq: ladder_lc}). The first of these results is the Klein-Gordon equation describing the oscillator in center-of-mass coordinates. The second is clearly a Lorentz covariant generalization of the non-relativistic oscillator eq. (\ref{eq: shrod2}) and the third is the constraint on the $\hat{a}_\mu^{\pm}$ operators reducing the number of independent components from four to three.

\section{Raising and Lowering Operations}
It is clear that each of the ladder operators $\hat{a}_\mu^{\pm}$ has four components but the three-dimensional harmonic oscillator only has three independent modes of oscillation.  In the rest frame of the oscillator, this is straightforward to interpret since eq. (\ref{eq: lo2}) gives $\hat{a}_4 \Psi = 0$ meaning there is one component of $\hat{a}_i^{\pm}$ for each spatial mode of oscillation. The complication for a moving oscillator is that $\hat{a}_4 \Psi$ does not vanish. The $\hat{P}^\mu \hat{a}_\mu^{\pm} \Psi = 0$ condition must therefore be interpreted as defining three linearly independent combinations of $\hat{a}_\mu^{\pm}$ components for the purpose of raising and lowering each of the three modes of oscillation. The task ahead is to determine the explicit form of these linear combinations.

A good starting point for the present task is to consider of the general Lorentz transformation equations:
\begin{equation} \label{eq: alt}
\hat{a}_i^{\pm \prime} = \hat{a}_i^{\pm} + \gamma v_i \left( \frac{\gamma v_j \hat{a}_j^{\pm}}{\gamma+1} + \hat{a}_4^\pm \right)
\end{equation} 
\begin{equation}
\hat{a}_4^{\pm \prime} = \gamma (\hat{a}_4^{\pm} +  v_i \hat{a}_i^{\pm}) = 0
\end{equation} 
that rotates the 4-vector $\hat{a}_\mu^{\pm \prime}$ from the rest frame of the oscillator to a moving frame where its components are denoted $\hat{a}_\mu^{\pm}$. Notice from eq. (\ref{eq: ladder_lc}) that eq. (\ref{eq: alt}) enables each of the components of $\hat{a}_i^{\pm \prime}$ defined in the rest frame of the oscillator to be expressed in terms of $\hat{x}_{\perp}^\mu$ and $\hat{p}^\mu$ operators in the moving frame. It is also helpful to spot that eq. (\ref{eq: alt}) can be simplified using eq. (\ref{eq: lo2}) to give
\begin{equation} 
\hat{a}_i^{\pm \prime} = \hat{a}_i^{\pm} - \frac{P_i \hat{a}_4^{\pm}}{(E+M_0)} 
\end{equation} 
having set $P_\mu = (P_i, E)$. This result can also be expressed in the form
\begin{equation} \label{eq: explicitladders}
\hat{a}_i^{\pm \prime} = \frac{1}{\sqrt{2 \Omega}}\left[ \mp \left( \frac{\partial}{\partial x_i} + \frac{P_i}{E+M_0} \frac{\partial}{\partial t} \right) +  \Omega x_i + \Omega\frac{P_i}{M_0} \left( \frac{P_jx_j}{M_0+E}-t\right) \right]
\end{equation} 
where the dependence on the $\hat{x}_{\perp}^\mu$ and $\hat{p}^\mu$ operators has been made explicit.

The explicit form of the oscillator function (\ref{eq: psi}) can be written as
\begin{equation} \label{eq: psi2}
\Phi=\frac{1}{\sqrt{2^n}} \left( \frac{\Omega}{\pi} \right)^{3/4} \Phi_1\Phi_2\Phi_3 
\end{equation} 
where
\begin{eqnarray} 
\Phi_i =\frac{1}{\sqrt{l_i!}} H_{l_i}\left[\sqrt{\Omega} x_i + \sqrt{\Omega}\frac{P_i}{M_0} \left( \frac{P_jx_j}{M_0+E}-t\right) \right] 
\nonumber \\
\exp \left\{-\frac{\Omega}{2} \left[x_i + \frac{P_i}{M_0} \left( \frac{P_jx_j}{M_0+E}-t\right) \right]^2 \right\}
\end{eqnarray}
Clearly, it can be seen
\begin{equation} \label{eq: simple1}
\left( \frac{\partial}{\partial x_i} + \frac{P_i}{E+M_0} \frac{\partial}{\partial t} \right) \left[x_j + \frac{P_j}{M_0} \left( \frac{P_kx_k}{M_0+E}-t\right) \right] = \delta_{ij}
\end{equation}
where $\delta_{ij}$ is the Kronecker delta. It follows from this result that
\begin{equation} \label{eq: simple2}
\frac{1}{\Phi}\left( \frac{\partial \Phi}{\partial x_i} + \frac{P_i}{E+M_0} \frac{\partial \Phi}{\partial t} \right) = \frac{1}{\Phi_i}\left( \frac{\partial \Phi_i}{\partial x_i} + \frac{P_i}{E+M_0} \frac{\partial \Phi_i}{\partial t} \right)
\end{equation}
Thus, applying eq. (\ref{eq: explicitladders}) to the oscillator function (\ref{eq: psi2}) it can be shown with the help of eqs. (\ref{eq: simple1}) and (\ref{eq: simple2}) that the relativistic ladder operators have the following set of properties.
\begin{equation} \label{eq: l1}
\hat{a}_1^{-\prime} \Phi(x_\mu,P_\mu,l_1, l_2, l_3) = \sqrt{l_1} \Phi(x_\mu,P_\mu,l_1-1, l_2, l_3)
\end{equation}
\begin{equation} \label{eq: l2}
\hat{a}_2^{-\prime} \Phi(x_\mu,P_\mu,l_1, l_2, l_3) = \sqrt{l_2} \Phi(x_\mu,P_\mu,l_1, l_2-1, l_3)
\end{equation}
\begin{equation} \label{eq: l3}
\hat{a}_3^{-\prime} \Phi(x_\mu,P_\mu,l_1, l_2, l_3) = \sqrt{l_3} \Phi(x_\mu,P_\mu,l_1, l_2, l_3-1)
\end{equation}
for lowering the quantum state of the oscillator, alongside 
\begin{equation} \label{eq: r1}
\hat{a}_1^{+\prime} \Phi(x_\mu,P_\mu,l_1, l_2, l_3)  = \sqrt{l_1+1} \Phi(x_\mu,P_\mu,l_1+1, l_2, l_3)
\end{equation}
\begin{equation} \label{eq: r2}
\hat{a}_2^{+\prime} \Phi(x_\mu,P_\mu,l_1, l_2, l_3)  = \sqrt{l_2+1} \Phi(x_\mu,P_\mu,l_1, l_2+1, l_3)
\end{equation}
\begin{equation} \label{eq: r3}
\hat{a}_3^{+\prime} \Phi(x_\mu,P_\mu,l_1, l_2, l_3)  = \sqrt{l_3+1} \Phi(x_\mu,P_\mu,l_1, l_2, l_3+1)
\end{equation}
for raising it. These fully relativistic expressions hold true in all inertial frames of reference.

Eqs. (\ref{eq: l1}) through (\ref{eq: r1}) can be used to show
\begin{equation} \label{eq: lo_prime}
\hat{a}_\mu^{+ \prime} \hat{a}^{\mu- \prime} \Phi = n \Phi
\end{equation}
Comparing eqs. (\ref{eq: lo1}) and (\ref{eq: lo_prime}) gives $\hat{a}_\mu^{+ } \hat{a}^{\mu-} = \hat{a}_\mu^{+ \prime} \hat{a}^{\mu- \prime}$ as is to be expected.

\section{Concluding Remarks}
It has been shown that the non-relativistic ladder operators $\hat{a}^{\pm}_i$, for raising and lowering the eigenstates of the harmonic oscillator, have a straightforward Lorentz covariant generalization $\hat{a}^{\pm}_\mu$ that reduces to the form $(\hat{a}^{\pm}_i,0)$ in the rest frame of the oscillator. Two significant results follow. Firstly, the $\hat{a}^{\pm}_\mu$ operators enable the quantum constraint equations for the relativistic oscillator to be expressed in a particularly concise form analogous to the simplification of the Schr\"{o}dinger equations for the harmonic oscillator in terms of $\hat{a}^{\pm}_i$ operators. Secondly, the constraint to three dimensions is shown to generate three independent linear combinations of $\hat{a}^{\pm}_\mu$ components for raising and lowering the eigenstates of the relativistic harmonic oscillator. The explicit forms of these linear combinations have been identified.

\end{document}